\documentclass[aps,prl,reprint,superscriptaddress]{revtex4-1}

\usepackage{textcomp}
\usepackage{upgreek}
\usepackage{array}
\usepackage{graphicx}
\usepackage{float}
\usepackage[none]{hyphenat}
\usepackage{amsmath}
\usepackage{epstopdf}

\begin{document}

\title{Crossed Andreev Reflection in InSb Flake Josephson Junctions}

\author{Folkert~K.~de~Vries}
\affiliation{QuTech and Kavli Institute of Nanoscience, Delft University of Technology, 2600 GA Delft, The Netherlands}
\author{Martijn~L.~Sol}
\affiliation{QuTech and Kavli Institute of Nanoscience, Delft University of Technology, 2600 GA Delft, The Netherlands}
\author{Sasa~Gazibegovic}
\affiliation{Department of Applied Physics, Eindhoven University of Technology, 5600 MB Eindhoven, The Netherlands}
\author{Roy~L.~M.~op~het~Veld}
\affiliation{Department of Applied Physics, Eindhoven University of Technology, 5600 MB Eindhoven, The Netherlands}
\author{Stijn~C.~Balk}
\affiliation{QuTech and Kavli Institute of Nanoscience, Delft University of Technology, 2600 GA Delft, The Netherlands}
\author{Diana~Car}
\affiliation{Department of Applied Physics, Eindhoven University of Technology, 5600 MB Eindhoven, The Netherlands}
\author{Erik~P.~A.~M.~Bakkers}
\affiliation{Department of Applied Physics, Eindhoven University of Technology, 5600 MB Eindhoven, The Netherlands}
\author{Leo~P.~Kouwenhoven}
\affiliation{QuTech and Kavli Institute of Nanoscience, Delft University of Technology, 2600 GA Delft, The Netherlands}	
\affiliation{Microsoft Quantum Lab Delft, 2600 GA Delft, The Netherlands}
\author{Jie~Shen}
\email{j.shen-1@tudelft.nl}
\affiliation{QuTech and Kavli Institute of Nanoscience, Delft University of Technology, 2600 GA Delft, The Netherlands}

\date{\today}
\begin{abstract}
We study superconducting quantum interference in InSb flake Josephson junctions.
An even-odd effect in the amplitude and periodicity of the superconducting quantum interference pattern is found.
Interestingly, the occurrence of this pattern coincides with enhanced conduction at both edges of the flake, as is deduced from measuring a SQUID pattern at reduced gate voltages. 
We identify the specific crystal facet of the edge with enhanced conduction, and confirm this by measuring multiple devices. 
Furthermore, we argue the even-odd effect is due to crossed Andreev reflection, a process where a Cooper pair splits up over the two edges and recombines at the opposite contact.
An entirely $h/e$ periodic SQUID pattern, as well as the observation of both even-odd and odd-even effects, corroborates this conclusion.
Crossed Andreev reflection could be harnessed for creating a topological state of matter or performing experiments on the non-local spin-entanglement of spatially separated Cooper pairs.
\end{abstract}

\maketitle
Induced superconductivity in semiconductors with strong spin-orbit interaction (SOI) attracted much interest for its potential applications in topological quantum computation~\cite{Alicea_2012}. 
A semiconducting Josephson junction (JJ) offers a platform to study the induced superconductivity by means of superconducting quantum interference (SQI)~\cite{Tinkham}.
Recently, induced superconductivity in edge channels in the quantum Hall regime~\cite{Amet_2016} and in a predicted two-dimensional topological insulator~\cite{Hart_2014,Pribiag_2015}, interesting for topological zero modes such as parafermions or Majoranas ,are investigated using SQI.
Additionally, an oscillation with both $h/e$ and $h/2e$ periodic components, before connected to topological edge states~\cite{Pribiag_2015}, is observed in a trivial InAs quantum well and attributed to crossed Andreev reflection (CAR) in the JJ~\cite{Baxevanis_2015,deVries_2018}.

Crossed Andreev reflection is a process where the quasiparticles that form a Cooper pair, are spatially separated but still entangled.
The entanglement of these quasiparticles holds promise in harnessing electrons in a solid-state environment to for example test the Einstein-Podolsky-Rosen paradox~\cite{EPR_1935} -- of fundamental importance to both quantum communication and computation.
Additionally, coupling two one-dimensional (1D) structures (i.e. nanowires or edge states) via CAR is interesting for engineering a topological state of matter hosting parafermions~\cite{Klinovaja_2014} or Majoranas\cite{Gaidamauskas_2014}.
To observe pronounced CAR in a device, normal or direct Andreev reflection needs to be suppressed.
In this regard, quantum dots~\cite{Recher_2001,Hofstetter_2009} or Luttinger liquids~\cite{Recher_2002} can be utilized.
Two-dimensional (2D) systems, such as a 2D electron gas~\cite{Russo_2005} connected to a superconductor, offer a scalable and flexible platform for more complex device geometries. 
Therefore, exploiting coupled 1D edge channels in a 2D material for Cooper pair splitting, combines the large CAR amplitude and flexibility in device design~\cite{Sato_2012,Baxevanis_2015}.

Here, we obtain measurements of CAR in a JJ made of an InSb flake -- a 2D nanostructure. 
We observe both even-odd and odd-even Fraunhofer patterns, and an entirely $h/e$ periodic SQUID pattern in JJs with enhanced conduction at both edges.
We argue that these $h/e$ effects are caused by a flux independent supercurrent due to CAR, where the quasiparticles are spatially separated over the two edges.

\begin{figure*}[t]
	\centering
	\includegraphics[width=6.2in]{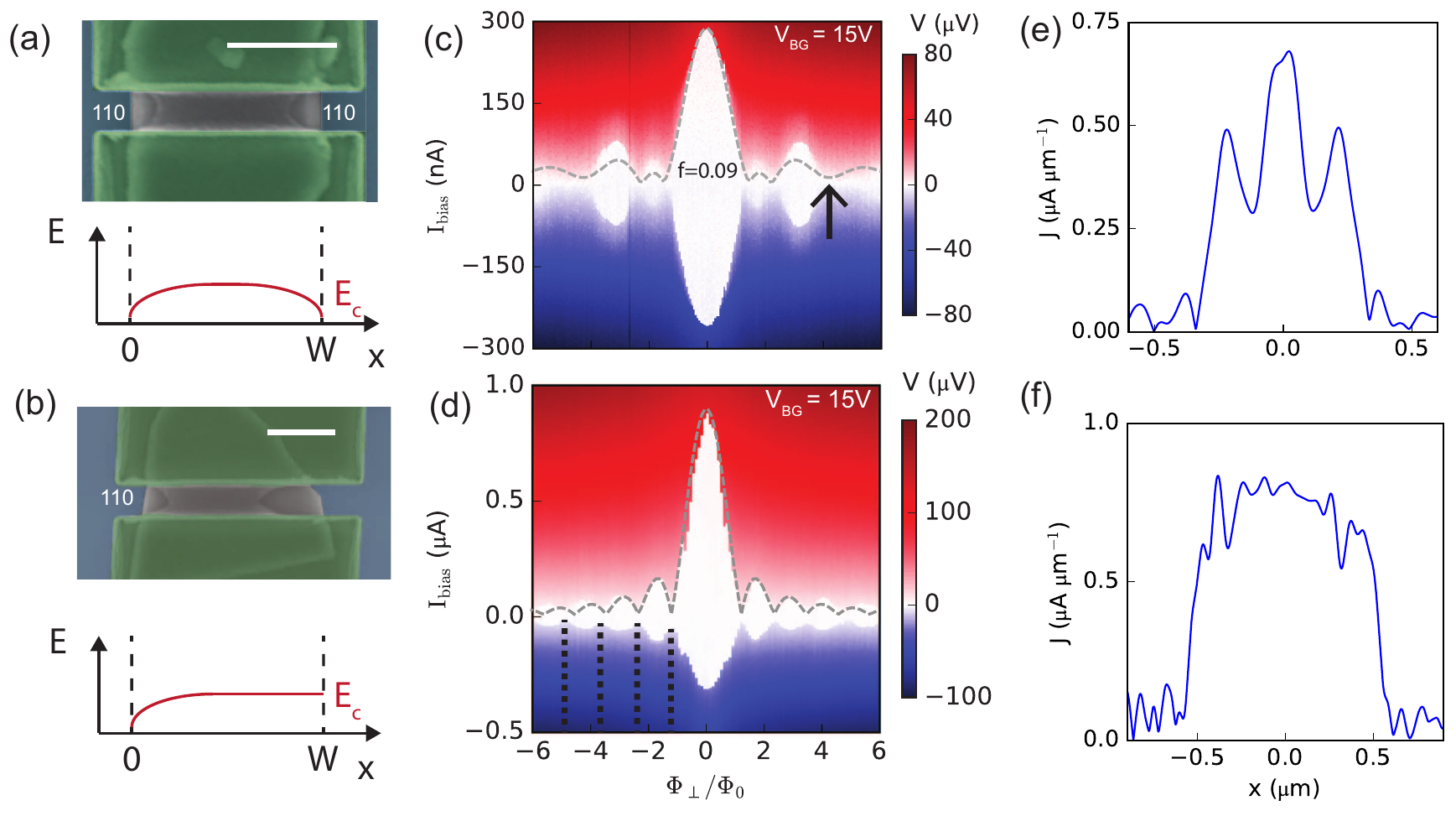}
	\caption{
(a,b) The upper panel shows a false colored scanning electron microscope image of JJ1 and JJ2, respectively. 
The flake (gray) is deposited on a Si/SiO$_\text{x}$ substrate (blue) and contacted by NbTiN (green). The scale bar represents 500\,nm. 
In the bottom panel, the energy of the bottom of the conduction band, $E_c$, (red) is sketched along the width, $W$, of the junction.
(c,d) Voltage, $V$, measured over JJ1 and JJ2 as a function of current bias, $I_\text{bias}$, and normalized flux, $\Phi/\Phi_0$ with $\Phi_0=h/2e$, through the JJ area, at a bottom gate voltage $V_\text{BG}$ of 15\,V. The dashed gray line is a calculated SQI pattern following Ref.~\citenum{Barzykin_1999} with an offset $f=0.09$ for (c) and no offset for (d)~\cite{SM}. For (c), the arrow highlights the missing third side lobe and in (d), the periodicity of the SQI pattern is indicated by the dashed black lines.
(e,f) Current density distribution, $J$, extracted from the SQI pattern of (c,d), using the Dynes-Fulton approach~\cite{Dynes_1971}.
	}
	\label{FF:fig1}
\end{figure*}

InSb is known for its large g-factor~\cite{delaMata_2016} and strong SOI~\cite{Nilsson_2009}, and earlier works referred to the flakes as nanosails~\cite{delaMata_2016} or nanosheets~\cite{Pan_2016,Zhao_2019}. 
The InSb flakes are grown with the vapor-liquid-solid technique~\cite{Gazibegovic_2017,SM}.
The crystal facet on their edges is (110) when stemming from a wire surface~\cite{SM}.
This (110) facet is known for having electron accumulation at its surface~\cite{Kreutz_1983,Ritz_1985}, because the lack of Sb atoms results in band bending~\cite{Whitman_1990}. 
Considering the geometry, we expect strong band bending at the edges of the flake with (110) facets as sketched in Fig.~\ref{FF:fig1}(a,b).
To fabricate devices, we use a micro-manipulator to transfer the flakes to a Si/SiO$_x$ substrate that serves as a global bottom gate.
Then, two NbTiN contacts are deposited after treating the surface with a sulfur solution to remove the native oxides~\cite{S_etching,Gul_2017}. 
The geometrical parameters, such as contact separation, $L$, and width, $W$, of all JJs are presented in the Supplemental Material~\cite{SM}. 
The JJs are measured in a quasi four-terminal current bias setup at a temperature of 300\,mK, unless stated otherwise.
Characterization of the superconductivity provides us an estimate of the superconducting gap, $\Delta$, of 1.4\,meV, consistent with values found earlier for NbTiN~\cite{Gul_2017}, and the induced superconducting coherence length, $\xi_s$, of 1.2\,\textmu m at $V_{\mathrm{BG}}=15$\,V (see Supplemental Material for details~\cite{SM}). 

\begin{figure}[t]
	\centering
	\includegraphics[width=\columnwidth]{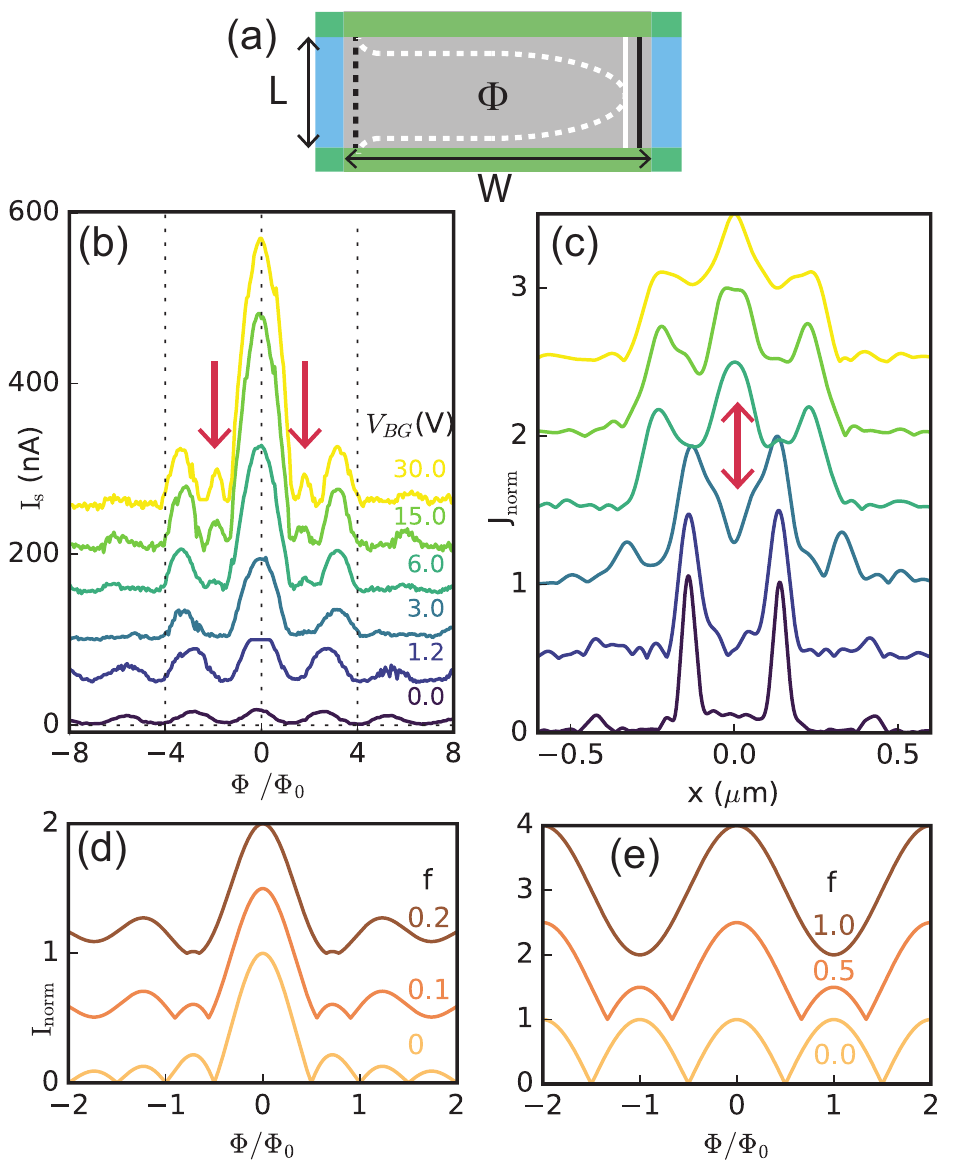}
	\caption{
(a) A sketch of crossed and direct Andreev reflection. The dashed (solid) line schematically represents a crossed (normal) Andreev reflection. The electron (white) encircles the junction, and picks up a phase due to flux, $\Phi$, while the hole (black) flows along the opposite edge.
(b) Switching current, $I_\text{s}$, as a function of normalized perpendicular magnetic flux, $\Phi/\Phi_0$, for JJ1 at the indicated gate voltages, $V_\text{BG}$. The red arrows highlight the first lobe, whose amplitude is diminished as $V_\text{BG}$ decreases.
(c) Normalized current density distributions, $J_\text{norm}$, extracted from the SQI patterns from (b). The disappearance of the central peak is highlighted by the red arrow.
(d,e) Calculated SQI patterns with a positive offset, $f$, for a standard Fraunhofer and SQUID pattern, respectively~\cite{SM}.
}
	\label{FF:fig2}
\end{figure}

Superconducting quantum interference measurements are performed by measuring the switching current of the JJs while varying the flux through them with an out-of-plane magnetic field.
Interestingly, the SQI pattern of JJ1 in Fig.~\ref{FF:fig1}(c) does not show the regular Fraunhofer pattern~\cite{Tinkham} as observed in Fig.~\ref{FF:fig1}(d) for JJ2.
The pattern instead displays an even-odd effect, which means the amplitude of the side lobes is not monotonically decaying but alternating.
The first side lobe has a smaller amplitude than the second, and the amplitude of the third side lobe is zero.
We describe this even-odd effect with a positive, magnetic field independent supercurrent offset $f$, added to the expected interference pattern $\mathcal{I}(\Phi)$:
\begin{equation}
\label{eq:SQI_f}
I_c(\Phi)=I_{c0}\left|\mathcal{I}(\Phi)+f\right| ,
\end{equation}
where $I_{c0}$ is the critical current at zero magnetic field, it increases (decreases) the amplitude of the lobes with the same (opposite) sign.
Examples of Fraunhofer and SQUID patterns with different positive offsets are presented in Fig.~\ref{FF:fig2}(d-e).
A SQUID pattern with such an offset is reported before~\cite{Pribiag_2015,deVries_2018}, however the even-odd Fraunhofer pattern is not experimentally studied to our knowledge.

\begin{figure}[t]
	\centering
	\includegraphics[width=\columnwidth]{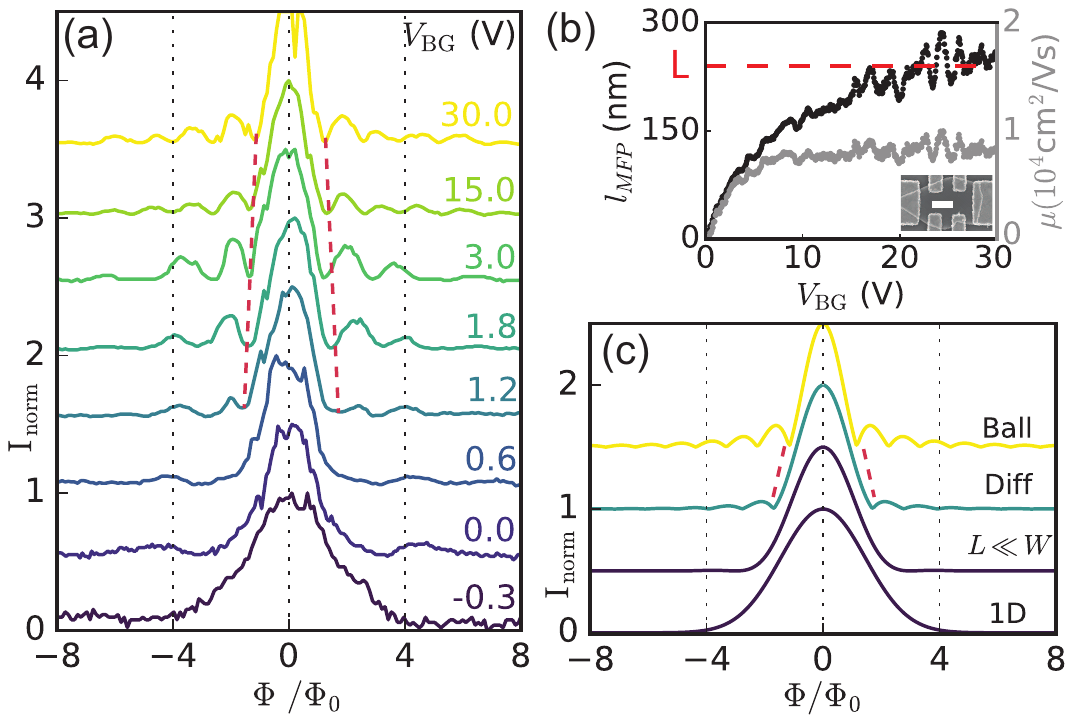}
	\caption{
(a) Normalized switching current, $I_\text{norm}$, as a function of normalized flux, $\Phi/\Phi_0$, for the indicated bottom gate voltages, $V_\text{BG}$. The dashed red lines highlight an increase in periodicity as $V_\text{BG}$ is reduced.
(b) Mean free path, $l_\text{MFP}$, and mobility, $\mu$, as a function of $V_\text{BG}$, extracted from Hall measurements~\cite{SM}. 
A SEM image of the Hall bar device is presented in the inset, with a scale bar representing 500\,nm. The length, $L$, of JJ2 is indicated by the dashed red line for comparison.
(c) The SQI pattern for a ballistic (Ball), diffusive (Diff) and quasi 1D ($L\gg W$) JJ are obtained from Ref.~\citenum{Barzykin_1999}, and for a 1D junction from Ref.~\citenum{Chiodi_2012}, of which details can be found in the Supplemental Material~\cite{SM}.
	}
	\label{FF:fig3}
\end{figure}

The SQI pattern in our InSb flake JJs cannot be described by the standard Fraunhofer pattern~\cite{Tinkham}, because our JJs does not satisfy the limit of $W\gg L$ (i.e. for JJ1 $W=1280$\,nm and $L=240$\,nm).
Therefore, we use a theoretical model for rectangular JJs as proposed by Barzykin~\textit{et al.} in Ref.~\citenum{Barzykin_1999}, where the supercurrent is calculated over all possible quasiparticle trajectories for either a ballistic or diffusive junction.
The expressions for the resulting SQI patterns for both cases are provided in the Supplemental Material for convenience~\cite{SM}.
We use the ballistic model in Fig.~\ref{FF:fig1} (dashed gray lines), since we estimate the mean free path, $l_\text{MFP}$, to be around 250\,nm, similar to $L$ [Fig.~\ref{FF:fig3}(b)].
The SQI pattern of JJ2 is well resembled by the calculated SQI pattern [Fig.~\ref{FF:fig1}(d)], showing that the larger periodicity of $1.5\Phi_0$ is due to the rectangular geometry of the JJ~\cite{Barzykin_1999}. 
Note that the SQI patterns are compensated already for flux focusing due to the Meissner effect~\cite{SM}.
In Fig.~\ref{FF:fig1}(c), the dashed gray line is calculated using the ballistic model~\cite{Barzykin_1999} with an offset of $f = 0.09$ incorporated with equation~\ref{eq:SQI_f}.
The even-odd behavior of the SQI pattern is qualitatively reproduced, supporting our assumption of a flux independent supercurrent.

In the calculations we have implicitly taken into account that the supercurrent is homogeneously distributed throughout the JJ.
To check this, we reconstruct the current density distribution with the method described by Dynes and Fulton~\cite{Dynes_1971}, which we are allowed to use, since the JJs are in the short junction limit, $\xi_s > L$~\cite{Hui_2014}.
The current density distribution for JJ2, plotted in Fig.~\ref{FF:fig1}(f), is homogeneously distributed, whereas for JJ1 [Fig.~\ref{FF:fig1}(e)] it reveals a large peak at the center of the JJ.
Apart from having a supercurrent through the center of the JJ, the peak could also be due to a magnetic field independent supercurrent. 
Because the Dynes-Fulton method is based on a Fourier transform, constant (or zero frequency) components end up at $x\,=\,0$ -- the center of the distribution.
Such a supercurrent offset cannot be due to a partial short in the JJ, since we confirmed the supercurrent can be pinched off by the global bottom gate~\cite{SM}.
Because of its insensitivity to magnetic field, the offset cannot stem from mechanisms that occur at a certain magnetic field either~\cite{Yokoyama_2014,Meier_2016}.
An effect that however could cause a magnetic field independent supercurrent is CAR~\cite{Recher_2001}.
CAR describes an Andreev pair of which one quasiparticle encircles the junction area [Fig.~\ref{FF:fig2}(a)] and therefore acquires a phase proportional to the flux through the junction area.
That extra phase can either directly~\cite{vanOstaay_2011} or by interference of two different Andreev pairs, result in a flux independent supercurrent~\cite{Baxevanis_2015,Jacquet_2015,Liu_2017}.

To find out whether a central current path or a magnetic field independent supercurrent due to CAR causes the even-odd effect, we continue by studying the gate dependence of the SQI patterns.
The even-odd SQI pattern from JJ1 changes drastically as a function of gate voltage [Fig.~\ref{FF:fig2}(b)]. 
The amplitude of the first side lobe decreases as $V_\text{BG}$ is reduced (highlighted by the red arrows), and becomes zero at $V_\text{BG}=3$\,V.
Then, for the bottom two traces of Fig.~\ref{FF:fig2}(b), the SQI pattern takes a cosinusoidal shape, known as a SQUID pattern~\cite{Tinkham}.
When the amplitude of the second lobe drops below the offset of the SQI pattern, the periodicity of the SQI pattern changes, see also the curve for $f=1$ in Fig.~\ref{FF:fig2}(e).
It doubles from 1.3$\Phi_0$ at $V_\text{BG}=15$\,V to 2.7$\Phi_0$ at $V_\text{BG}=3$\,V, and the SQI pattern becomes entirely $h/e$ periodic for $V_\text{BG}\leq1.2$\,V.
This is different from the observed $h/e$ SQUID in Ref.~\citenum{Pribiag_2015} and Ref.~\citenum{deVries_2018}, where the amplitude was larger than the offset and therefore an $h/2e$ oscillation is observed simultaneously.
Our observation confirms that the $h/e$ periodicity is not a unique signature of a topological JJ~\cite{deVries_2018}.

The changes in the SQI pattern are reflected in the extracted current density distributions in Fig.~\ref{FF:fig2}(c).
Note that $J$ is spanning half of the width for $V_\text{BG}\leq3$\,V compared to $V_\text{BG}\geq6$\,V, because we used the same area and flux periodicity for the calculation of all traces.
The center peak in the current density disappears at $V_\text{BG}=3$\,V.
Such a local effect in $J$ is not likely to be caused by changing the global gate.
Additionally, the offset in the SQI patterns persists, even though there is no center peak anymore.
Therefore, we disregard a current path at the center of the JJ as an explanation for the even-odd effect.
Interestingly, between $V_\text{BG}=3$\,V and 0\,V, the SQUID pattern translates to a current density distribution with edge conduction only [bottom trace of Fig.~\ref{FF:fig2}(c)], in agreement with JJ1 having electron accumulation at both edges [Fig.~\ref{FF:fig1}(a)].
This enhanced conduction at both edges of the JJ, in combination with the $h/e$ periodicity is consistent with the occurrence of CAR~\cite{Baxevanis_2015,deVries_2018}.
To substantiate this, we consider this mechanism in detail, and study additional JJs.

The CAR trajectories encircling the JJ area consist of the edges of the flake, and two paths along the contacts, as sketched in Fig.~\ref{FF:fig2}(a).
The latter could arise of combining doping from and a finite barrier to the superconducting contact~\cite{deVries_2018}.
The estimated induced superconducting coherence length of 1.2\,\textmu m is close to the typical junction circumference of 2\,\textmu m~\cite{SM}.
Due to the difference between the circumference and $L$, the CAR is expected to be suppressed with temperature before direct Andreev reflection (and supercurrent) diminishes. 
A temperature dependence is however ambiguous, since the side lobes (and with that the even-odd effect) disappear before the switching current is suppressed~\cite{SM}.
The magnitude of the offset reaches a value of 1 in Fig.~\ref{FF:fig2}(b), which is in range of what one can expect for a combination of a small coupling to the contact, while maintaining the Fermi velocity in the edges~\cite{Baxevanis_2015,SM}.
Additionally, electron-electron interaction in the 1D edges could also reduces the direct Andreev reflection and lead to a large f~\cite{Sato_2019}.
The the large offset $f$ and $h/e$ periodicity mean that the CAR amplitude exceeds the direct Andreev reflection, an interesting topic and regime for future experiments.

To shine light on the correlation between the even-odd effect and having enhanced conduction at both edges, we study the gate dependence of JJ2 as well.
The periodicity of the SQI patterns in Fig.~\ref{FF:fig3}(a) grows slightly from $1.3\Phi_0$ to $1.5\Phi_0$ as $V_\text{BG}$ is lowered from 30\,V to 1.2\,V.
Meanwhile the mobility $\mu$ and $l_\text{MFP}$ decrease [Fig.~\ref{FF:fig3}(b)], and for $V_\text{BG} <3$\,V the length $L$ of the JJ is larger than $l_\text{MFP}$ and the JJ changes from ballistic to diffusive.
In Fig.~\ref{FF:fig3}(c), calculated SQI traces for ballistic and diffusive transport are plotted \cite{Barzykin_1999}, consistent with the transition observed in our data. 
Furthermore, the strong increase in periodicity and suppression of the side lobe amplitudes for gate voltages below $V_{\mathrm{BG}} =1.2$\,V highlight a transition from a 2D to a 1D diffusive regime.
The measured SQI pattern at $V_\text{BG}=-0.3$\,V is well reproduced by the theoritical curves for a (quasi) 1D JJ [Fig.~\ref{FF:fig3}(c)], as described~\cite{Barzykin_1999,Chiodi_2012} and observed before~\cite{Chiodi_2012,Amado_2013}. 
Entering the 1D regime is in line with having a single edge with (110) facet and enhanced edge conduction in JJ2 [Fig.~\ref{FF:fig1}(b)].
Continuing the argument, not finding an even-odd effect in JJ2, strengthens the connection between the even-odd SQI pattern and having enhanced conduction at both edges.

\begin{figure}[t]
	\centering
	\includegraphics[width=\columnwidth]{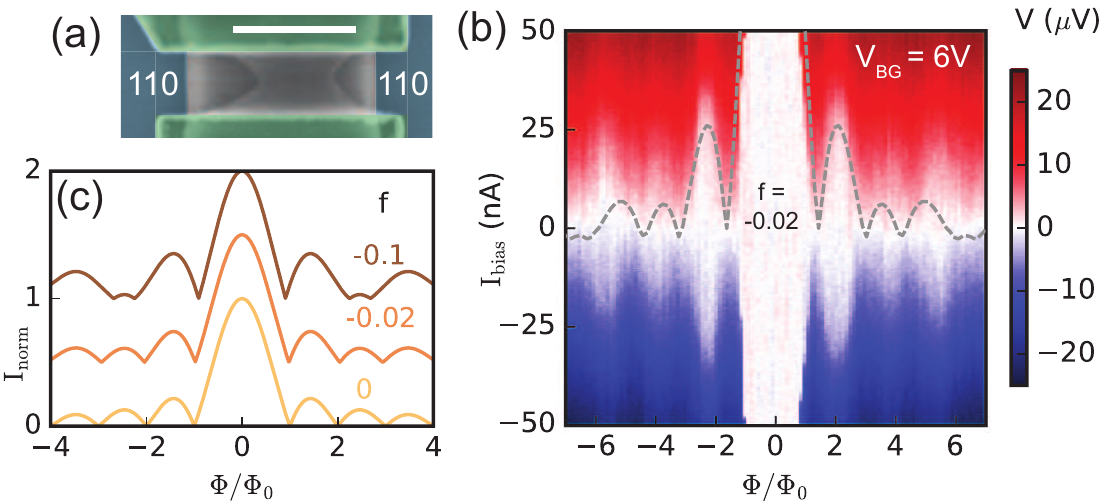}
	\caption{
(a) False colored SEM image of JJ3. The flake (gray) is deposited on a Si/SiO$_\text{x}$ substrate (blue) and contacted by NbTiN (green). The scale bar represents 500\,nm.
(b) SQI pattern of JJ3 at $V_\text{BG} = 15$\,V. The dashed gray line is a simulation following Ref.~\citenum{Barzykin_1999} with offset $f=-0.02$~\cite{SM}.
(c) Calculated Fraunhofer patterns with a variable (negative) offset, $f$, as indicated.
}
	\label{FF:fig4}
\end{figure}

Two other devices (JJ3 and JJ4) also reveal an even-odd SQI pattern, and show enhanced conduction in both their edges with a (110) facet~\cite{SM}.
Furthermore, JJ5-JJ7, having a single (110) edge, do not show an even-odd SQI pattern~\cite{SM}.
Interestingly, the SQI pattern from JJ3 [Fig.~\ref{FF:fig4}(b)] shows a negative offset, or odd-even effect. 
The third lobe has a similar amplitude to the second, thus is not smaller as expected for a standard Fraunhofer pattern.
By adding a negative offset to the calculated SQI pattern from Ref.~\citenum{Barzykin_1999} [dashed gray line in Fig.\ref{FF:fig4}(b)], we indeed find good agreement with the data.
For comparison we plotted the standard Fraunhofer patterns with negative offsets in Fig.~\ref{FF:fig4}(c).
Having either a positive or negative offset to the switching current due to CAR depends on microscopic details regarding the spin mixing in the JJ~\cite{Baxevanis_2015}.
To be more precise, spin mixing with predominantly spin conserving or spin-flip processes, refer to a positive or negative offset, respectively. 
In our InSb flakes, the spin mixing is probably caused by strong spin-orbit interaction in the InSb~\cite{Nilsson_2009}. 
The observation of a negative offset is, to our knowledge, unique to CAR~\cite{Baxevanis_2015}, and therefore strongly supports that CAR is causing the observed $h/e$ periodic SQI patterns.
	
In conclusion, we observe $h/e$ effects due to crossed Andreev reflection in the SQI patterns of InSb flake Josephson junctions with enhanced conduction at both edges.
We identified crystal facet (110) to have enhanced edge conduction, and can thus in the future choose to either study or circumvent them.
The observed $h/e$ SQUID pattern reveals that the CAR amplitude can exceed the direct Andreev reflection in a 2D semiconducting Josephson junction.
The InSb flakes therefore provide a promising platform to use CAR for creating topological zero modes~\cite{Klinovaja_2014,Gaidamauskas_2014} or for applications in Cooper pair splitting~\cite{EPR_1935,Sato_2012}. 


\section*{Acknowledgments}
This work has been supported by funding from the Netherlands Organization for Scientific Research (NWO) and Microsoft.

\section*{Author contributions}
S.C.B developed the substrate preparation, S.G. grew the flakes, D.C. and E.P.A.M.B. contributed to the growth and R.L.M.o.h.V developed the flake deposition procedure.
J.S., F.K.d.V. and M.S. fabricated the devices, performed the measurements and analyzed the data. 
F.K.d.V. and J.S. wrote the manuscript with comments from all authors.
J.S. and L.P.K. supervised the project.


%

\onecolumngrid
\linespread{1.5}

\setcounter{figure}{0}

\makeatletter 
\renewcommand{\thetable}{S\arabic{table}}%
\renewcommand{\thefigure}{S\arabic{figure}}%
\renewcommand\figurename{\normalsize \textbf{Figure}}
\renewcommand\tablename{\normalsize\textbf{Table}}
\newpage

\begin{center}
\qquad \\[0.1cm] \LARGE{ \huge Supplemental Material \\[5mm] \LARGE Crossed Andreev Reflection in InSb Flake Josephon Junctions}\\[1cm]

\linespread{1.5}
\large \noindent Folkert~K.~de~Vries, Martijn~L.~Sol, Sasa~Gazibegovic, Roy~L.~M.~op~het~Veld, Stijn~C.~Balk, Diana~Car, Erik~P.~A.~M.~Bakkers, Leo~P.~Kouwenhoven and Jie~Shen\\[2cm]
\end{center}

\normalsize
\clearpage

\section{InSb flakes growth}
\begin{figure}[!ht]
	\centering
	\includegraphics[width=.9\columnwidth]{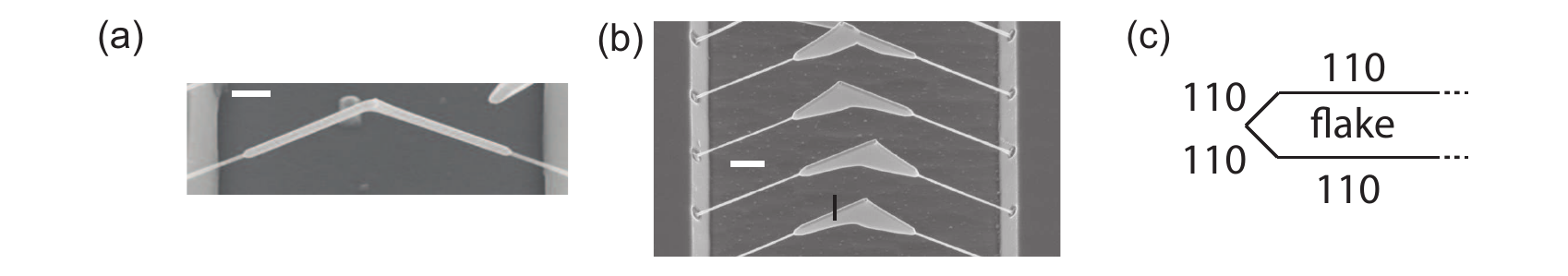}
	\caption{
(a-b) A scanning electron microscope (SEM) image of the InSb flakes, grown from two nanowires on an InP substrate~\cite{Gazibegovic_2017_SM}. 
The scale bar represents 500\,nm. 
(a) The etched trenches in an InP substrate allow for growing coalescing nanowires. 
(b) After two nanowires touch, the growth continues in the two-dimensional plane spanned by them.
(c) Sketch of the cross section of the flake, at the location highlighted by the solid black line in (b).
All facets, top, bottom and edge have (110) crystal facets. Note that these are the edge facets stemming from the nanowire.
}
	\label{FF_SM:fig1}
\end{figure}

\newpage
\section{Devices}
\begin{table}[h]
  \begin{tabular}{ l || c | c | c | c | c | c | c  }
   &\textbf{JJ1} & \textbf{JJ2} & \textbf{JJ3} &\textbf{JJ4} & \textbf{JJ5} & \textbf{JJ6} & \textbf{JJ7} \\ \hline
   $W$ (nm)					&770& 1280  & 855 & 790 & 1360 & 970 &  1150\\ \hline
   $L$ (nm) 					&240& 240  & 185 & 425 & 205 & 465 &  670\\ \hline
   $L/W$ 						&0.31& 0.19  & 0.22 & 0.54 & 0.15 & 0.48 & 0.58 \\ \hline
   $A$ ($\mu$m$^2$)			&0.18& 0.31  & 0.16 & 0.34 & 0.28 & 0.45 & 0.77\\ \hline
   $A_{\mathrm{eff}}$ ($\mu$m$^2$)		&0.63& 0.74  & 0.95 & 1.21 & 1.37 & 0.81 & 1.03\\
\end{tabular}
\caption{Geometry parameters for all 7 JJs. The width, $W$, and length, $L$, are obtained from the SEM images. The areas, $A$ and $A_\mathrm{eff}$, are the relevent areas for the applied perpendicular magnetic field, where the effective area includes flux focusing from the superconducting contacts due to the Meissner effect.}
\label{FF_SM:tab1}
\end{table}
\begin{figure}[h]
	\centering
	\includegraphics[width=\columnwidth]{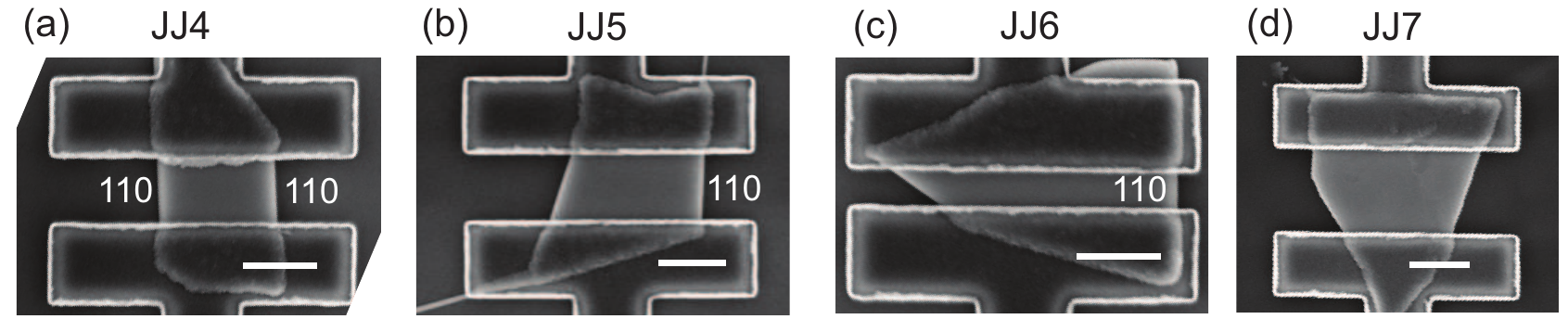}
	\caption{SEM images of JJ4-JJ7, (a-d), respectively. The scale bar represents 500 nm. Note that JJ4 in (a) has parallel edges, both with (110) crystal faceting.
	}
	\label{FF_SM:fig2}
\end{figure}

\newpage
\section{Induced superconductivity}
\begin{figure}[H]
	\centering
	\includegraphics[width=\columnwidth]{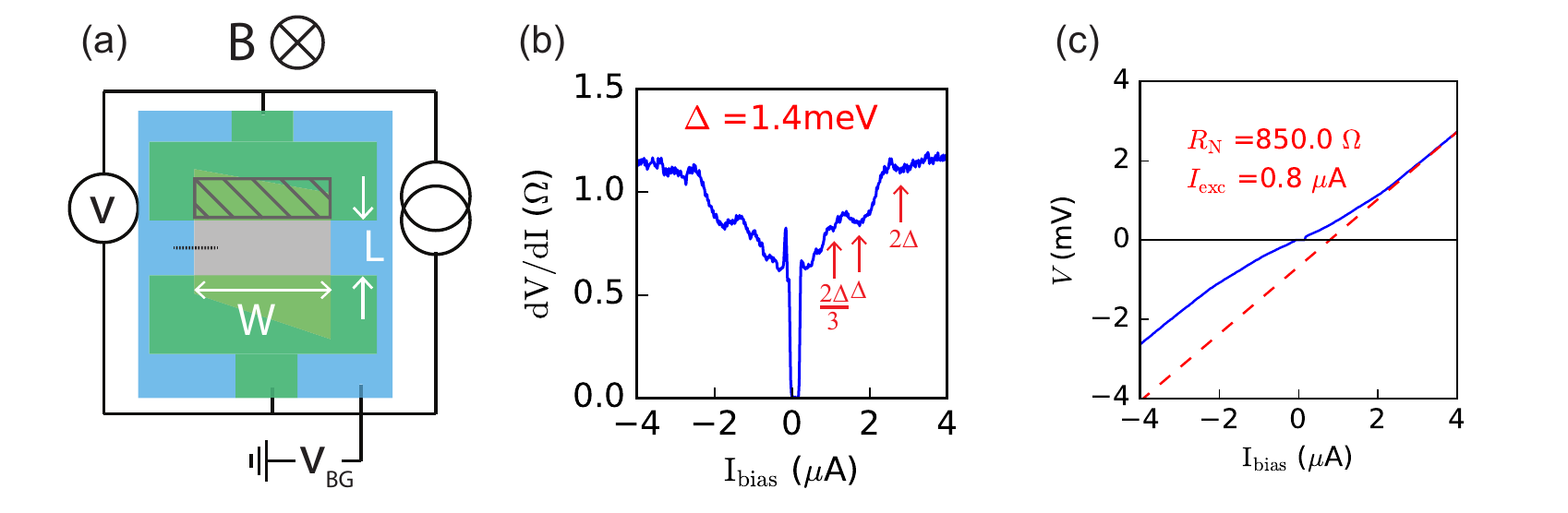}
	\caption{(a) Schematic Josephson junction device, where the flake (gray) is contacted by NbTiN (green), after being placed on the Si/SiO substrate (blue). 
The superconducting contacts focus the flux of the hatched area into the junction, due to the Meissner effect.
The four terminal current bias measurement setup is indicated as well as the magnetic field, $B$. 
(b) Differential resistance, $dV/dI$, measurement on JJ2 as a function of current bias, $I_\text{bias}$, at a bottom gate voltage, $V_\text{BG}$, of 2.4\,V. 
A superconducting gap, $\Delta$, of 1.4\,meV is extracted from the resonances at twice the gap edge and the multiple Andreev reflections, indicated by the red arrows. 
(c) Voltage measured, $V$, of JJ2 as a function of $I_\text{bias}$ at $V_\text{BG}=2.4$\,V. The excess current, $I_\text{exc}$, and normal state resistance, $R_\text{N}$, are extracted and used to estimate a transmission of 0.6, following Ref.~\citenum{Flensberg_1988_SM}.
	}
	\label{FF_SM:fig3}
\end{figure}
\begin{figure}[H]
	\centering
	\includegraphics[width=\columnwidth]{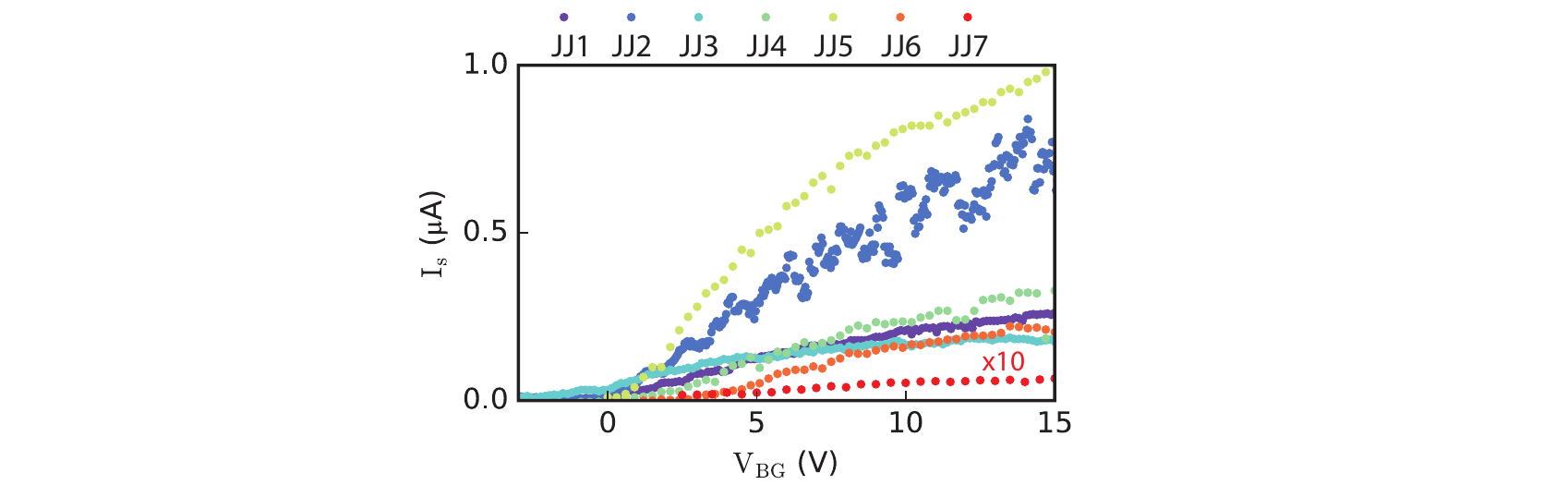}
	\caption{Switching current $I_\mathrm{s}$ as a function of bottom gate voltage $V_\mathrm{BG}$ for all JJs.}
	\label{FF_SM:fig4}
\end{figure}

\section{Superconducting quantum interference models}
\subsection{Fraunhofer and SQUID SQI patterns}
The standard Fraunhofer and SQUID SQI patterns are given by \cite{Tinkham_SM}:
\begin{equation}
I_\text{c} (\Phi)=I_\text{c,0}\left|\frac{\sin(\pi \Phi/\Phi_0)}{\pi \Phi/\Phi_0}\right|,
\end{equation}
\begin{equation}
I_\text{c} (\Phi)=I_\text{c,0}\left|\cos(\pi \Phi/\Phi_0)\right|,
\end{equation}
where $I_\text{c,0}$, is the critical current at zero flux, $\Phi$ is the flux and $\Phi_0=h/2e$ the superconducting flux quantum.

\subsection{Ballistic and difussive SQI}
The superconducting quantum interference in a JJ with a length, $L$, and finite width, $W$, is described analytically in Ref.~\cite{Barzykin_1999_SM}.
For convenience we included their main results here.
The final expressions for the critical current as a function of flux in a ballistic JJ ($I_\text{c,ball}$) and diffusive JJ ($I_\text{c,diff}$) are \cite{Barzykin_1999_SM}:
\begin{equation}
\label{FF:eqballistic}
\begin{aligned}
I_\text{c,ball} (\Phi)=&\max_{0\leq\phi\leq2\pi} \frac{2ev_\text{F}}{\pi W\lambda_\text{F} L}\iint^{W/2}_{-W/2} \frac{dx_1dx_2}{\left[1+\left(\frac{x_1-x_2}{L}\right)^2\right]^{3/2}} \\
&\sum_{k=1}^\infty (-1)^{k+1} \frac{L}{\xi_\text{T} \cos \theta_{x_1-x_2} }\frac{\sin k \left[\frac{\pi \Phi}{W \Phi_0}(x_1+x_2)+\phi\right]}{\sinh \frac{kL}{\xi_\text{T} \cos \theta_{x_1-x_2}}};
\end{aligned}
\end{equation}
\begin{equation}
\label{FF:eqdiffusive}
\begin{aligned}
I_\text{c,diff}(\Phi) &\propto\sum^\infty_{l=-\infty} (-1)^l S_l(L/2) \frac{d}{du}S_l(u)\Bigr|_{\substack{u=L/2}} \left(\frac{\sin[\pi(\Phi+l)/2]}{\pi(\Phi+l)/2} - (-1)^l \frac{\sin[\pi(\Phi-l)/2]}{\pi(\Phi-l)/2}\right)^2, \\ 
S_l(u)&=\sqrt{|u|/2\pi}(q_\text{T}^2+\pi^2l^2/W^2)^{1/4} K_{1/2}\left(\sqrt{u^2(q_\text{T}^2+\pi^2l^2/W^2)}\right)
\end{aligned}
\end{equation}\\
where, $\phi$ is the superconducting phase difference, $v_\text{F}$ and $\lambda_\text{F}$ are the Fermi velocity and wavelength, $\xi_\text{T}$ is the normal metal coherence length ($\hbar v_\text{F} / 2\pi k_\text{B} T$), $\tan \theta_{x_1-x_2}=(x_2-x_1)/L$ is the angle of the Andreev reflection with respect to the vector perpendicular to the contact.
$K_{1/2}$ is a modified Bessel equation and $q_\text{T}=1/\tilde{\xi}_\text{T}^2$, where $\tilde{\xi}_\text{T}$ is the diffusive coherence length.
The SQI pattern in the limit of $L \ll W$ for a diffusive JJ reads \cite{Barzykin_1999_SM}:
\begin{equation}
f_\text{diff}(\nu)=\frac{\cos^2\pi \nu/2}{(1-\nu^{2})^{2}}.
\end{equation}

The geometrical parameters used for the different devices, can be found in table~\ref{FF_SM:tab1}.
Furthermore, we use an effective mass of 0.02~$m_e$ \cite{Qu_2016_SM} and a temperature of 300\,mK (or 50\,mK) to extract parameters from the Hall bar measurement (Fig.~\ref{FF_SM:fig6}).
For the calculated SQI patterns in Fig.~1 and Fig.~4, at $V_\text{BG}=15$\,V, we find $v_\text{F}=2.5\cdot 10^6$ \,m/s, $\lambda_\text{F}$=\,2.3\,nm, and $\xi_\text{T}$=10 $\mu$m; and for Fig.~3, at $V_\text{BG}=3$\,V, $v_\text{F}=1.5\cdot 10^6$ \,m/s, $\lambda_\text{F}$=\,4.0\,nm, and $\xi_\text{T}$=5.9 $\mu$m.
Because the devices are in the crossover regime from ballistic to diffusive transport, we use the ballistic coherence length in both cases, so we set $\tilde{\xi_\text{T}}=\xi_\text{T}$.
With a transmission value of $T\leq0.6$ we expect to have a sinusoidal current phase relation \cite{Haberkorn_1978_SM}, and therefore only use $k=1$ for eq.~\ref{FF:eqballistic}. 
For eq.~\ref{FF:eqdiffusive} we sum up to $l=100$. For both we confirmed that the periodicity does not change for summing over a larger $k$ or $l$, respectively.
In all figures in the main text we plot the normalized critical current, either $I_\text{c}(\Phi)/I_\text{c}(0)$ or $f_\text{diff}(\Phi)/f_\text{diff}(0)$.

\subsection{1D diffusive model}
For a one-dimensional system in the diffusive limit, the critical current decays monotonically, following Ref.~\citenum{Chiodi_2012_SM}:
\begin{equation}
\label{eq:1dlimit}
I_c (\Phi)\propto f=e^{-0.238\Phi^2}.
\end{equation}

\subsection{CAR offset $f$}
The offset in the SQI pattern caused by CAR is expressed as~\cite{Baxevanis_2015_SM}:
\begin{equation}
f \sim \Gamma^{-1} \frac{k_B T}{\Delta} e^{-2\pi (k_B T / \Delta) (W / \xi_s)}.
\end{equation}
In example for JJ1: $T=300$\,mK, $W\,=\,770$\,nm, $\Delta=1.4$\,meV, $\xi_s=1.2$\,\textmu m; resulting in $f=\Gamma^{-1}\cdot0.02$.
To reach $f=1$, $\Gamma=0.02$, which means there should be a relatively strong barrier between the contacts and the edge states.

\newpage
\section{Temperature dependence SQI pattern JJ1}
\begin{figure}[h]
	\centering
	\includegraphics[width=\columnwidth]{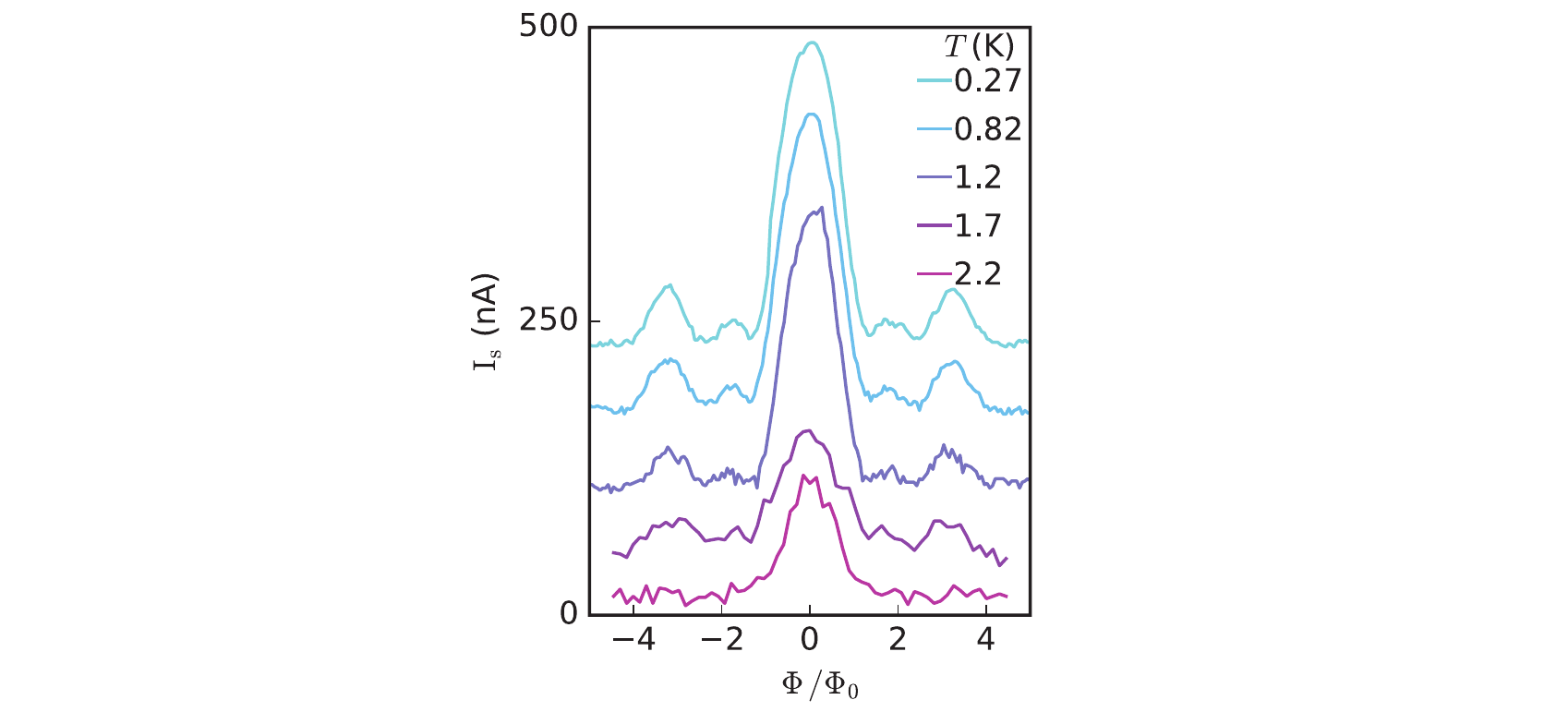}
	\caption{Temperature dependence of the SQI pattern of JJ1 at $V_{BG}=15$\,V. For crossed Andreev reflection we do expect the even-odd effect to die out earlier than the supercurrent, because the coherence length needed is the junction circumference that is larger than the junction length. The side lobes (and with that the even-odd effect) diminishes before the switching current is suppressed, which makes the temperature dependence inconclusive. 
	}
	\label{FF_SM:fig5}
\end{figure}

\newpage
\section{Hall bar measurement}
\begin{figure}[h]
	\centering
	\includegraphics[width=\columnwidth]{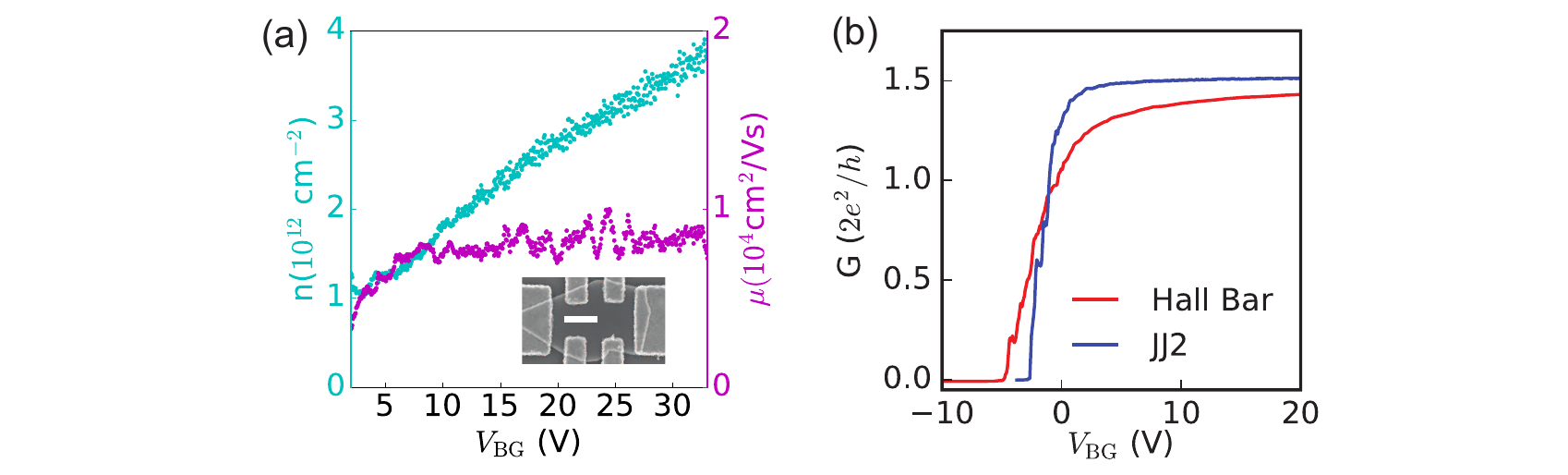}
	\caption{(a) Hall bar measurements are used to extract the density, $n$, and mobility, $\mu$, as a function of bottom gate voltage, $V_\mathrm{BG}$.
	The Hall bar device (inset) is fabricated following the same steps as for the JJs, except it has Cr/Au metal contacts. The scale bar represents 500\,nm.
	The density is converted to fermi velocity, $v_\text{F}$, and then used to estimate the induced superconducting coherence length, $\xi_s=h v_\text{F}/\Delta$, with $\Delta=1.4$\,meV as obtained from Fig.~\ref{FF_SM:fig3}.
	(b) Conductance, measured at a constant DC voltage bias, as a function of $V_\mathrm{BG}$ for both the Hall bar device and JJ2. The saturation is reflecting the sum of contact and fridge line resistances. Comparing both pinch off curves suggests that the density (and with that $v_\text{F}$) is larger in JJ2. This means qualitatively that we are underestimating $\xi_s$.
	}
	\label{FF_SM:fig6}
\end{figure}

\newpage
\section{Extra devices: JJ3-JJ7}

\begin{figure}[h]
	\centering
	\includegraphics[width=\columnwidth]{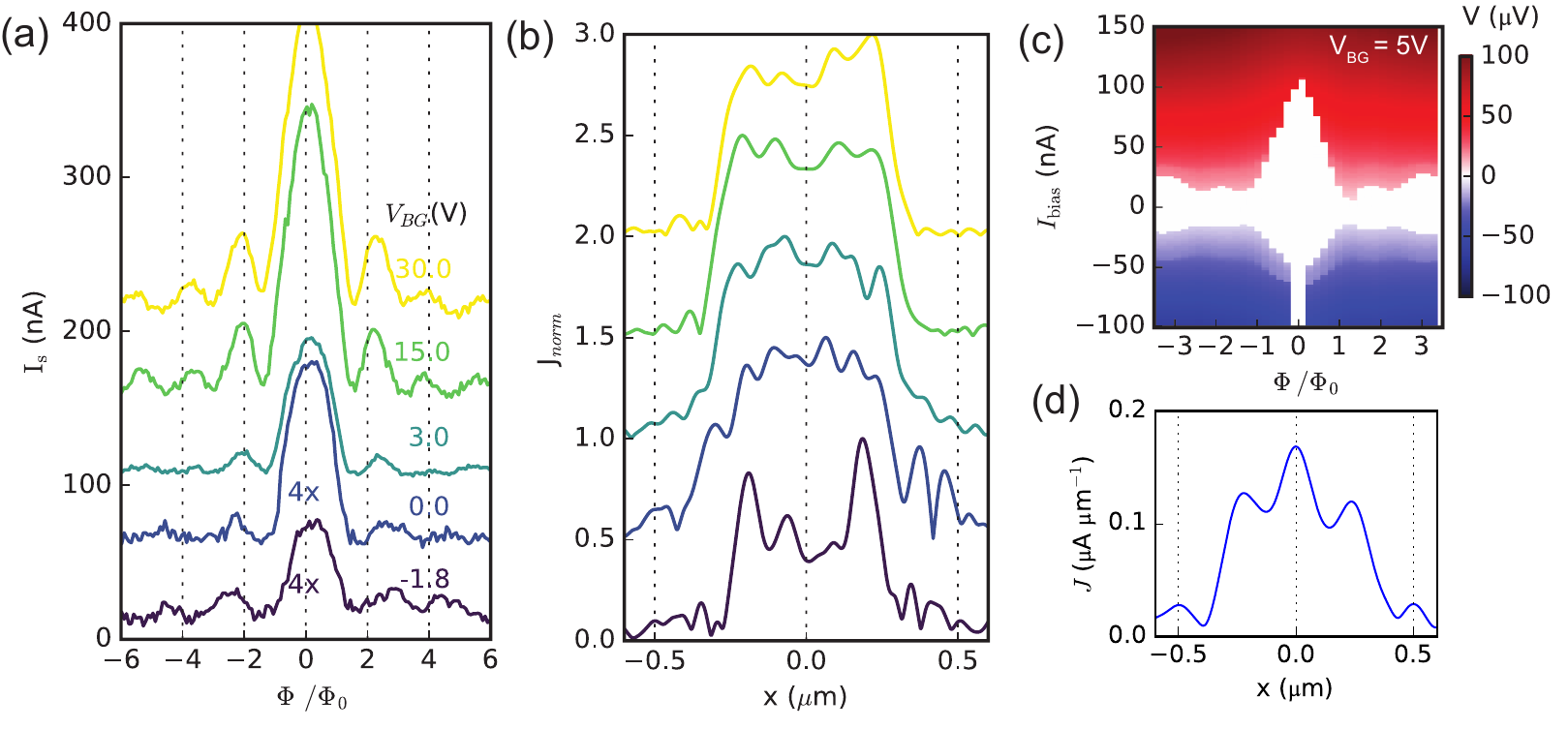}
	\caption{
(a) Switching current, $I_s$, of JJ3 as a function of normalized flux, $\Phi/\Phi_0$, at the indicated gate voltages. The bottom two SQI patterns are multiplied by 4 for visibility.
(b) Normalized current density distribution, $J_\mathrm{norm}$, calculated from the patterns in (a). 
At $V_\text{BG}=-1.8$\,V the enhanced edge conduction shows up.
(c) An even-dd SQI pattern is observed for JJ4. Note that JJ4 has two parallel edges with (110) faceting. This measurement is performed at a temperature of 50\,mK.
(d) $J$ calculated from the SQI pattern in (c). The even-odd effect is reflected in the central peak in $J$. 
	}
	\label{FF_SM:fig7}
\end{figure}

\begin{figure}[h]
	\centering
	\includegraphics[width=\columnwidth]{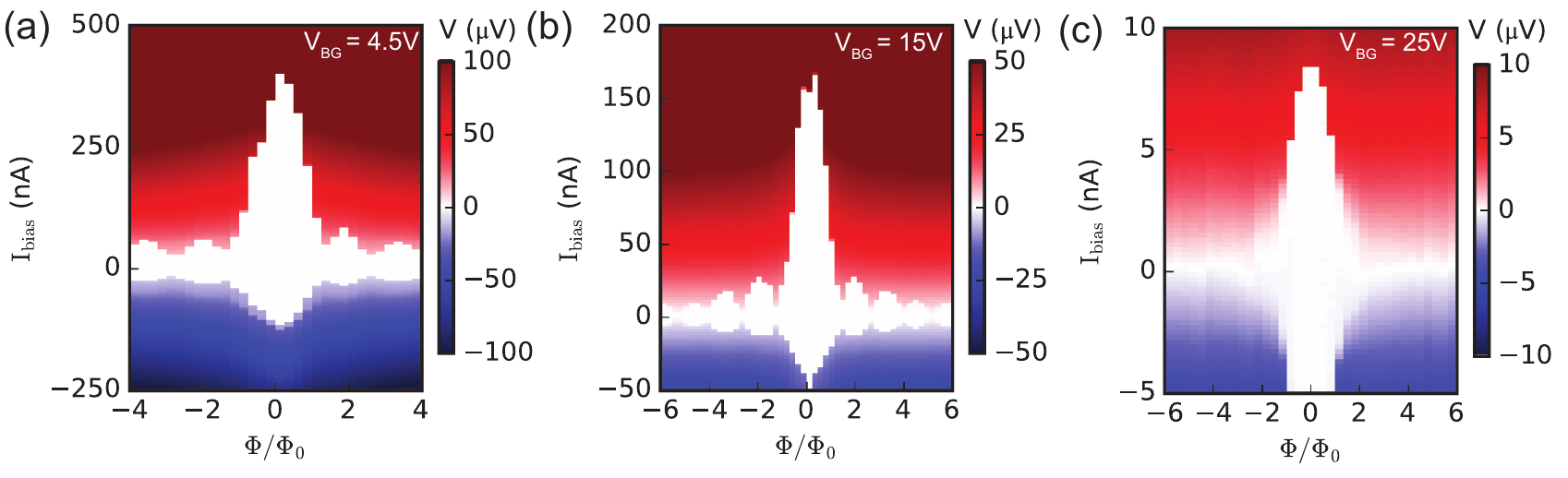}
	\caption{(a-c) SQI patterns of JJ5-JJ7, respectively, at gate voltages indicated. For (a,b) we observe a standard Fraunhofer pattern, as expected since both JJs do not have two edges with (110) faceting. In (c) the Fraunhofer pattern for JJ7 is different because of the larger length ($L=1.15$\,\textmu m) of this device. It resembles the interference pattern of a 1D diffusive junction, likely reflecting a narrow path contributing only to the supercurrent. All measurements are performed at $T=50$\,mK.
}

	\label{FF_SM:fig8}
\end{figure}

\clearpage
\newpage

\end{document}